\documentclass[showkeys,reprint,nofootinbib]{revtex4-1}
\usepackage{graphicx}% Include figure files
\usepackage{times}
\usepackage{mathrsfs}
\usepackage{amsmath,amssymb}

\allowdisplaybreaks
\newcommand{\eqhyp}[1]{#1\nobreak\discretionary{}{\hbox{\ensuremath{#1}}}{}}

\def \arccot {\mathop{\rm arccot}\nolimits}

\begin{document}
\abovedisplayskip=3pt plus 3.0pt minus 3.0pt
\abovedisplayshortskip=0.0pt plus 3.0pt
\belowdisplayskip=3pt plus 3.0pt minus 3.0pt
\belowdisplayshortskip=3pt plus 3pt minus 3pt

\title{\LARGE Analog of the Auger effect in radiative decay of a trion in~a~quantum well}
\author{N.A. Poklonski}
\email{poklonski@bsu.by}
%\thanks{}
\author{A.I. Syaglo}
\author{S.A. Vyrko}
\affiliation{Belarusian State University, pr. Nezavisimosti 4, 220030 Minsk, Belarus}


\date{\today}

\begin{abstract}
We have analyzed the energetics of decay of the $X^-$ trion (exciton + electron) on the assumption that the exciton and trion are independent excitations of a single two-dimensional semiconducting quantum well. For the first time, it has been shown that in filling a well with electrons from a selective donor-doped matrix, the binding energy of the trion (of the electron with the exciton) increases linearly with a shift of the Fermi level into the depth of the \emph{c} band. This agrees with the well-known experimental data on the low-temperature radiative decay (photoluminescence) of trions in the heterostructures ZnSe$/$Zn$_{0.89}$Mg$_{0.11}$S$_{0.18}$Se$_{0.82}$ and CdTe$/$Cd$_{0.7}$Mg$_{0.3}$Te.
\end{abstract}

%Use showkeys class option if keyword display desired

\maketitle %\maketitle must follow title, authors, abstract and \pacs

%\section{Introduction}

In [1, 2], radiative decay of trions (negative ions of excitons [3]) in a single quantum well of width $L_z = 8$~nm in the semiconductor structures ZnSe$/$Zn$_{0.89}$Mg$_{0.11}$S$_{0.18}$Se$_{0.82}$ and CdTe$/$Cd$_{0.7}$Mg$_{0.3}$Te was investigated at a temperature of 1.6~K by the methods of magnetooptics. Conduction electrons ``were supplied'' to the quantum well (see Fig. 1) from a chlorine-doped 3~nm-thick $\delta$-layer located in the matrix at a distance of 10~nm from the quantum well [2]. Studies of the reflection and photoluminescence spectra of the quantum wells of ZnSe and CdTe have revealed that the intensity of the trionic line and the binding energy of the trion in the wells with a larger concentration of conduction electrons is higher, whereas that of the exciton is lower, than in the case of wells with a smaller concentration of electrons. The increase in the binding energy of the trion\footnote{The calculation of the dependence of the ground state energy of the $X^-$ trion (exciton + electron) and of the $X^+$ trion (exciton + hole) in a two-dimensional quantum well as functions of the ratio between the masses of the electron and hole is given in [4].} with increase in the concentration of conduction electrons is explained [1, 2] by the interaction of exciton and trion excitations of the well (having captured an additional electron, the $X$ exciton is converted into the $X^-$ trion, while the trion that lost the electron is converted into the exciton), i.e., by jumps of the electron from the trion onto the exciton. However, quantitative evaluations by the above model were not presented.

\begin{figure}%[!h]
\noindent\includegraphics[width=\columnwidth]{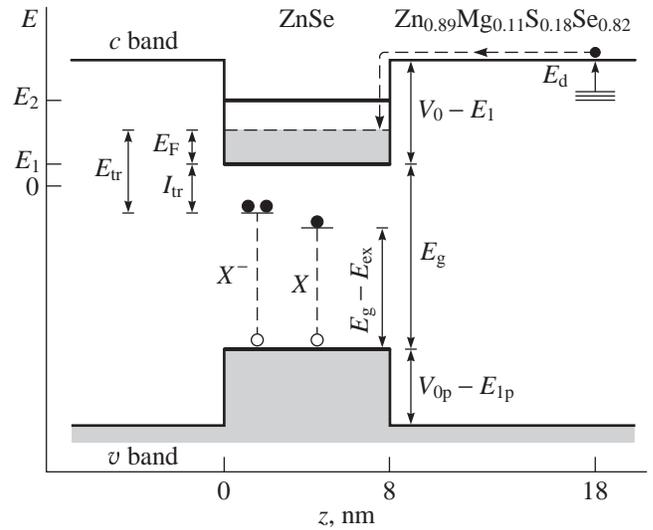}
\caption{Band diagram of the single ZnSe quantum well investigated in [1, 2]. Open circles are holes; solid circles are electrons; $X$ is an exciton, $X^-$ is a trion; $E_\text{g} \approx 2840$~meV is the forbidden-band width; $E_\text{ex}$ is the binding energy of the exciton; $E_\text{tr}$ is the energy of separation of the electron from the $X^-$ trion; $I_\text{tr}$ is the binding energy of one electron in a trion for the \emph{c} band of the quantum well barely filled with electrons ($E_\text{F} = 0$); $E_\text{d}$ is the mean ionization energy of the donor (Cl atom) in the matrix.}\label{fig:01}
\end{figure}

We will analyze the luminescence of trions and excitons using as an example the quantum well of ZnSe. The forbidden-band width of the considered quantum well at $L_z = 8$~nm with account for dimensional quantization [5] is $E_\text{g} \approx 2840$~meV. According to [2], the rupture of the \emph{c}~band for this structure is $V_0 - E_1 \approx 100$~meV, which is larger than the ionization energy of donors (Cl atoms in the Zn$_{0.89}$Mg$_{0.11}$S$_{0.18}$Se$_{0.82}$ matrix). In the ZnSe single crystal, the energy of the detachment of an ``optical'' electron from a chlorine atom is $E_\text{d} \approx 26$~meV [6], so that the two-dimensional concentration of the conduction electrons $n$ in the quantum well is equal to that of the completely ionized hydrogen-like donor impurity $N_{+1}$ in the matrix.

According to [1], with increase in the concentration of two-dimensional electrons from $n = 10^9$~cm$^{-2}$ to $5{\cdot}10^{11}$~cm$^{-2}$, the binding energy of the trion $E_\text{tr}$ (the binding energy of the electron with exciton) increases from 4.5 to 6.3~meV for the ZnSe well, and from 2.7 to 5.1~meV for the CdTe well. At the same time, as the concentration of the two-dimensional conduction electrons increases, the binding energy of the two-dimensional exciton $E_\text{ex}$ decreases.

According to [7], the decrease in the binding energy of the exciton $E_\text{ex}$ with increase in $n$ can be associated with screening of the Coulomb interaction between the hole with an electron and the filling of the pulse space with two-dimensional electrons. With increase in the concentration of electrons in the quantum well, the photoluminescence line that corresponds to the trion decay is then shifted to the long-wave region, i.e., the trion binding energy is increased. This is in contradiction with the idea that the screening by conduction electrons must break a trion [1--3].

We suggest explaining the increase [1] in the binding energy of a trion with increase in the concentration of conduction electrons in a single quantum well as an analog of the Auger process\footnote{The Auger effect (process) is a three-particle process which is manifested in recombination of an electron of the \emph{c} band with a hole of the $v$ band when all the energy (or a portion of it) liberated in this case is transferred to another electron of the \emph{c} band [8--11].} that is manifested in radiative decay of the trion. We carry out a quantitative evaluation of this effect, proceeding from the energy conservation law for the particles participating in recombination with account for a partial filling of the states of the \emph{c} band in the quantum well with conduction electrons.

At the temperature $T \to 0$, the binding energy of a trion $E_\text{tr}$ in the quantum well is the difference between the energies of the finite state (of the exciton with total energy $E_X$ and of the electron in the \emph{c} band with kinetic energy $E_\text{F}$ counted from the first quantum level $E_1$) and the initial state (of the trion with total energy $E_{X^-}$):
\begin{equation}\label{eq:01}
   E_\text{tr} = E_X + E_\text{F} - E_{X^-} = I_\text{tr} + E_\text{F},
\end{equation}
where $I_\text{tr} = E_X - E_{X^-}$ is the binding energy of a trion in relation to disintegration into an exciton and an electron with zero kinetic energy ($E_\text{F} = 0$) and the total energy $E_1$, i.e., for a vanishingly small concentration of conduction electrons in the quantum well.

In formula (1), it was taken into account that the kinetic energy of the electron that passed into the \emph{c} band of the quantum well on disintegration of a trion into an exciton and an electron cannot be smaller than $E_\text{F}$, since at $T = 0$ all the states below the Fermi level are occupied by electrons. As the temperature rises, the quantity $E_\text{tr}$ decreases because of the smearing of the boundary between the occupied and free states of the \emph{c} band.

Theoretical evaluations of the ratio $I_\text{tr}/I_\text{ex}$, where $I_\text{ex}$ is the binding energy of an exciton $E_\text{ex}$ at the zero concentration of two-dimensional electrons, are given in [4].

Let us consider the energetics of the decay of a trion (a bound state of a hole and two electrons) in a quantum well in much the same way as was done in [12] for the ionization energy of hydrogen-like impurities in a crystalline semiconductor depending on their concentration (see also [13]). On total decay of a trion, one electron goes into the \emph{c} band\footnote{An analogy with the superconductivity of metals [14]: the formation of Cooper pairs (of an energy gap $\Delta$) leads to a situation where the conduction electron requires the additional energy  to occupy the free state above the Fermi surface. The part of the Cooper pairs in the quantum well is played by the layer of states (filled with electrons) between the first quantum level $E_1$ and the Fermi level $E_\text{F}$.}, with its kinetic energy being equal to the Fermi energy $E_\text{F}$ or larger, while another electron recombines with a hole of the $v$ band emitting a photon with energy $\hbar\omega_\text{tr}$. With account for expression (1), this gives
\begin{equation}\label{eq:02}
   \hbar\omega_\text{tr} \approx \hbar\omega_\text{ex} - E_\text{tr} = \hbar\omega_\text{ex} - I_\text{tr} - E_\text{F},
\end{equation}
where $E_\text{g} - E_\text{ex} = \hbar\omega_\text{ex}$ is the photon energy liberated in radiative recombination of an exciton with the binding energy $E_\text{ex}$ in a quantum well; $E_\text{tr} \approx \hbar\omega_\text{ex} - \hbar\omega_\text{tr}$ is the binding energy of a trion.

From relations (1) and (2) it follows that with increase in the concentration of conduction electrons in a quantum well (against a neutralizing background of positively charged donors), the Fermi energy $E_\text{F} > 0$ increases, as a result of which $E_\text{tr}$ increases, while $\hbar\omega_\text{tr}$ decreases. This can be interpreted as an analog of the Auger process [8--11]: a portion of the energy released as a result of recombination of the electron-hole pair is transferred to the electron that goes into the unoccupied state of the \emph{c} band with kinetic energy $E_\text{F}$. At the same time, the radiative decay of a trion is also similar to the Burstein--Moss shift.\footnote{The Burstein--Moss shift [8--11] consists of the increase in energy needed for excitation of an electron from the $v$ band to the \emph{c} band during the filling of the states of the \emph{c} band due to the strong doping of the semiconductor. Here, as the concentration of conduction electrons grows, the edge of the optical self-absorption is shifted toward the high energies by the same quantity as the Fermi level into the \emph{c} band.}

Now, we evaluate the energy of the transverse motion of electrons and the Fermi level $E_\text{F}$ in the quantum well presented in Fig.~1.

The energy of the $j$th quantum-dimensional level $E_j$ of the transverse motion of an electron with effective mass $m$ in the quantum well of depth $V_0$ and width $L_z$ can be found from the following transcendental equation [15--17]:
\begin{equation}\label{eq:03}
   E_j = \biggl(\frac{\pi\hbar}{L_z}\biggr)^2 \frac{1}{2m} \Biggl[j - \frac{2}{\pi} \arccot \sqrt{\frac{m}{m_\text{b}}\biggl(\frac{V_0}{E_j} - 1\biggr)}\Biggr]^2\!\!,
\end{equation}
where $m_\text{b}$ is the effective mass of the electron in the matrix; $\hbar = h/2\pi$ is the Planck constant.

Figure 1 presents the calculated (from Eq.~(3)) levels $E_1 \eqhyp\approx 21$~meV and $E_2 \approx 78$~meV that were counted from the bottom of the quantum ZnSe well of depth $V_0 \approx 121$~meV for $m_\text{b} \approx m \approx 0.15m_0$; $E_{1\text{p}} \approx 7$~meV is the first quantum level of the transverse motion of a heavy hole with effective mass $m_\text{p} \approx 0.6m_0$ in the well $V_{0\text{p}} \approx 107$~meV.

When only the first quantum level is filled, the coupling between the two-dimensional concentration of electrons $n$ in the quantum well and the Fermi level $E_\text{F}$ for the temperature $T \to 0$ has the form [17--19]
\begin{equation}\label{eq:04}
   n = \frac{m}{\pi\hbar^2}E_\text{F}.
\end{equation}

We note that formulas (3) and (4) are obtained on the assumption of the local electrical neutrality, i.e., when both the electrons and ionized donors with the two-dimensional concentration $N_{+1} = n$ are located in the quantum well. In accordance with [18], in calculating the Fermi level $E_\text{F}$ in the quantum well for $T \to 0$, we can restrict ourselves to the filling of only the first quantum level ($j = 1$) with the conduction electrons, when $n < 3\pi/2L_z^2 = 7.4{\cdot}10^{12}$~cm$^{-2}$ for $L_z = 8$~nm.

The ground state of the trion is singlet [20], since the two electrons that compose the trion have oppositely directed spins. As a consequence, according to [1, 2], in a strong enough magnetic field $B$, when all the conduction electrons in the quantum well are spin-polarized, the trion can be manifested only in one circular polarization of the photons which interact with it. The degree of the circular polarization of the trion line in the reflection spectra has a maximum at an odd number of the filled Zeeman components of the Landau levels [11]:
\begin{equation}\label{eq:05}
   \nu = 2\alpha + 1 = \frac{2mE_\text{F}}{\hbar eB} = n\frac{2\pi\hbar}{eB},
\end{equation}
where $\alpha = 0, 1, \ldots$ is the number of the Landau level of the conduction electrons in the quantum well, which is located in the perpendicular magnetic field. With an even number $\nu \eqhyp= 2(\alpha + 1)$ of filled Landau sublevels, the circular polarization degree of the trion line is minimum.

Figure 2 presents the dependence of the change in the binding energy of a trion $\hbar\omega_\text{ex} - \hbar\omega_\text{tr} - I_\text{tr}$ on the position of the Fermi level (concentration of conduction electrons) in quantum wells. It should be noted that the values of the Fermi level $E_\text{F}$ plotted on the abscissa axis were determined from the experimental data on the polarization degree of the trion lines in the reflection and absorption spectra by using formulas (4) and (5) [1, 2] (Fig.~2).

\begin{figure}[!t]
\noindent\includegraphics%[width=0.75\columnwidth]
{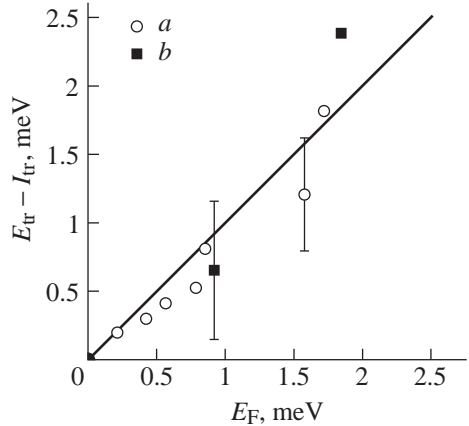}
\caption{Dependence of the change in the trion binding energy on the position of the Fermi level $E_\text{F}$ counted from the first quantum level $E_1$ in the \emph{c} band for the ZnSe and CdTe quantum wells at $T \eqhyp= 1.6$~K. Experimental data [1]: \emph{a})~ZnSe$/$Zn$_{0.89}$Mg$_{0.11}$S$_{0.18}$Se$_{0.82}$, \emph{b})~CdTe$/$Cd$_{0.7}$Mg$_{0.3}$Te. Line, the values of $E_\text{tr} - I_\text{tr} = E_\text{F}$ calculated at $T \to 0$ from formula (1) with $I_\text{tr} = 4.5$ and $2.7$~meV for \emph{a} and \emph{b}, respectively.}\label{fig:02}
\end{figure}

Expressions (4) and (5) can be applied only in the case where the electric field between the quantum well and the $\delta$-layer of donors in the matrix does not influence the energetics of optical processes in the quantum well. The electric field strength $F$ between the quantum well and the $\delta$-layer (see Fig.~1) can be evaluated from the formula for a plane capacitor $F = en/\varepsilon_\text{r}\varepsilon_0$, where $-en < 0$ is the two-dimensional density of the negative charge in the quantum well equal to the density of the positive charge of the donors $+eN_{+1} > 0$ in the $\delta$-layer of the matrix; $\varepsilon_\text{r} = 7.6$. Then, as the electron concentration of electrons $n$ in the ZnSe well increases from $10^9$ to $5{\cdot}10^{11}$~cm$^{-2}$, the internal electric field $F$ changes from 240 to $1.2{\cdot}10^5$~W$/$cm. Since this field is confined in the main beyond the quantum well and the electron density is shifted to the well wall (toward the $\delta$-layer of the donors), this is equivalent to a certain decrease in the well width $L_z$.

Thus, proceeding from the energy conservation law and an analogy with the Auger effect, for the first time a quantitative description has been given for the increase in the binding energy and for the shift in the trion photoluminescence line to the low-energy region on increase in the concentration of conduction electrons in a single quantum well. The consideration is carried out for $T \to 0$, i.e., in the limit of the zero energy of the translational motion of the trion and exciton in the quantum well plane.

%\begin{acknowledgments}
%\end{acknowledgments}

%\bibliography{poklonski}

\def\bibsection{\bigskip\textbf{\refname}}

\end{document}